\documentclass{emulateapj}

\usepackage{natbib}
\usepackage{rotating}
\usepackage{epsfig}
\usepackage{psfig}
\usepackage{amsmath,amssymb}

\newcommand\deltakv{\delta_{\bf k}}

\newcommand\deltaminuskv{\delta_{\bf -k}}
\newcommand\kcutoff{k_{\rm cutoff}}
\newcommand\kv{{\bf k}}
\newcommand\kzero{k_0}
\newcommand\lzero{l_0}

\renewcommand\i{{\rm i}}
\newcommand\Ie{I_{\rm e}}
\newcommand\Nx{N_x}
\newcommand\Ny{N_y}
\newcommand\Nz{N_z}
\newcommand\Pzero{P_0}
\newcommand\qx{q_x}
\newcommand\qy{q_y}
\newcommand\qz{q_z}
\newcommand\rzero{r_0}
\newcommand\rhalf{r_{\rm half}}
\newcommand\rhozero{\rho_0}
\newcommand\rhobar{\bar{\rho}}
\newcommand\rhohalf{\rho_{\rm half}}

\newcommand\rs{r_{\rm s}}
\newcommand\rhobg{\rho_{\rm bg}}
\renewcommand\Re{R_{\rm e}}

\newcommand\tu{t_{\rm u}}
\newcommand\tdyn{t_{\rm dyn}}
\newcommand\xv{{\bf x}}

\newcommand\vx{v_x}
\newcommand\vy{v_y}
\newcommand\vz{v_z}
\def\FVFPS{{\sc fvfps}}
\def\FORTRAN{{\sc fortran}}
\def\Sersic{{S\'ersic}}

\defcitealias{Agu90}{AM90}
\defcitealias{Kat91}{K91}

\journalinfo{The Astrophysical Journal Letters, accepted 2015 May 4}
\submitted{Accepted 2015 May 4} 

\shorttitle{Gaussian random field and S\'ersic law}
\shortauthors{Nipoti}

\begin{document}


\title{Gaussian random field power spectrum and the S\'ersic law}


\author{Carlo Nipoti}
\affil{Department of Physics and Astronomy, Bologna University, viale Berti-Pichat 6/2, 40127 Bologna, Italy}
\email{e-mail: carlo.nipoti@unibo.it}


\begin{abstract}
The surface-brightness profiles of galaxies are well described by the
{\Sersic} law: systems with high {\Sersic} index $m$ have steep central
profiles and shallow outer profiles, while systems with low $m$ have
shallow central profiles and steep outer profiles. R. Cen (2014, ApJL,
790, L24) has conjectured that these profiles arise naturally in the
standard cosmological model with initial density fluctuations
represented by a Gaussian random field (GRF). {We explore and
  confirm this hypothesis with $N$-body simulations of dissipationless
  collapses in which the initial conditions are generated from GRFs
  with different power spectra.} The numerical results show that GRFs
with more power on small scales lead to systems with higher $m$. In
our purely dissipationless simulations the {\Sersic} index is in the
range $2\lesssim m\lesssim 6.5$. {It follows that systems with
  {\Sersic} index as low as $m \approx 2$ can be produced by coherent
  dissipationless collapse, while high-$m$ systems can be obtained if
  the assembly history is characterized by several mergers.}  As
expected, dissipative processes appear to be required to obtain
exponential profiles ($m\approx 1$).
\end{abstract}





\section{Introduction}

The surface-brightness profiles of spheroidal and disk components of
galaxies are well described by the \citet{Ser68} law
\begin{equation}
I(R)=\Ie \, \exp\left\{-b(m)\left[ \left( \frac{R}{\Re} \right)^{1/m}
  -1 \right]\right\},
\label{eq:ser}
\end{equation}
where $\Re$ is the effective radius, $m$ is the {\Sersic} index,
$\Ie\equiv I(\Re)$ is the surface brightness at the effective radius
and $b(m)\simeq 2m-1/3+4/(405m)$ \citep{Cio99}.  When $m$ is high the
central profile is steep and the outer profile is shallow, while when
$m$ is low the central profile is shallow and the outer profile is
steep. Recently \citet{Cen14} has envisaged that these profiles arise
naturally in the standard cosmological model with initial density
fluctuations represented by a Gaussian random field (GRF). The
underlying idea is that central concentrations of stars and extended
envelopes are formed by the late infall and accretion of
substructures. Therefore, if the fluctuation field is dominated by
long-wavelength modes the formation of the galaxy, due to the absence
of significant substructures, is mainly determined by a coherent
collapse, and the final profile will be shallow in the center and
steep in the outskirts (an extreme case is the exponential profile
$m=1$, which is typical for disks). If the fluctuation field is
dominated by short-wavelength modes there is substantial late infall
of substructures and the final profile will be steeply rising toward
the center and gently declining in the outskirts (as observed, for
instance, in massive elliptical galaxies with $m\gtrsim4$).

In this Letter we explore quantitatively Cen's proposal with numerical
experiments in which we follow the dissipationless collapse of cold
distributions of particles whose initial conditions are determined by
GRFs with different power spectra.  Numerical simulations of
dissipationless collapse have been run by several authors
\citep[][]{van82,Udr93,Boi02,Nip06,Joy09,Syl13,Ben15,Wor15}. {A
  general finding is that the end-products have surface-density
  profiles well fitted by the {\Sersic} law (equation~\ref{eq:ser}).}
Remarkably, this result is not specific to Newtonian gravity, as it is
also found in studies of dissipationless collapses in modified gravity
theories \citep{Nip07,Dic13}. In Newtonian gravity the final {\Sersic}
index is typically close to $m=4$ (\citealt{van82,Agu90}, hereafter
\citetalias{Agu90}), the value corresponding to the \citet{dev48}
profile. Though there are some indications that the final value of $m$
can depend on the initial conditions \citep{Tre05,Nip06}, so far there
is no clear evidence of a dependence of $m$ on the properties of the
initial fluctuations.  Since the seminal work of \citet{van82} it was
realized that the clumpiness of the initial conditions is an important
factor in determining the nature of the collapse end-product.  Clumpy
initial conditions were considered in several investigations
\citep{May84,McG84,Lon91,Roy04,Tre05}, but in these works the
inhomogeneities of the initial phase-space distribution were not
systematically classified in terms of fluctuation power spectrum.
{An exception is \citet[][hereafter \citetalias{Kat91}]{Kat91},
  who set up the initial conditions self-consistently from a GRF and
  explored different power spectra (see also \citealt{Dub91} and
  \citealt{Bin08}, section 4.10.3).}

Here we present high-resolution $N$-body simulations aimed at
isolating the effect of the spectrum of the inhomogeneities of the
initial conditions on the final density profiles of cold
dissipationless collapses. We present evidence that the {\Sersic} index
of the collapse end-product correlates with the slope of the
fluctuation power spectrum of the initial conditions.

\section{Numerical experiments}

\begin{table}
\caption{Properties of the simulations. $\Pzero$ and $n$: amplitude
  and power-law index of the fluctuation power-spectrum of the initial
  conditions . $c/a$ and $b/a$: final shortest-to-longest and
  intermediate-to-longest axis ratios. $\rhalf/\rzero$: final
  half-mass radius in units of the scale radius $\rzero$. $m$ and
  $\sigma_m$: best-fitting {\Sersic} index of the final density
  profile and associated uncertainty.\label{tab:par}}
\begin{tabular}{lrrrrrrr}
\tableline
\tableline
\noalign{\vskip 2mm}   
Name  & $\Pzero$  & $n$ &  $c/a$ & $b/a$ & $\rhalf/\rzero$ & $m$ & $\sigma_m$ \\
\tableline
\noalign{\vskip 2mm}   
 P0 & $0$ & $-$     & 0.43 & 0.56 & 0.92 & 2.01 & 0.04\\
 P03n3 & $0.3$ & -3 & 0.46 & 0.71 & 1.05 & 2.26 & 0.02\\
 P03n25 & $0.3$ & -2.5 & 0.46 & 0.76 & 1.04 & 2.23 & 0.02\\
 P03n2 & $0.3$ & -2 & 0.48 & 0.83 & 1.08 & 2.38 & 0.02\\
 P03n15 & $0.3$ & -1.5 & 0.53 & 0.91 & 1.05 & 2.61 & 0.04\\
 P03n1 & $0.3$ & -1 & 0.61 & 0.98 & 1.01 & 3.32 & 0.12\\
 P03n05 & $0.3$ & -0.5 & 0.60 & 0.93 & 0.68 & 4.90 & 0.15\\
 P03n0 & $0.3$ & 0 & 0.63 & 0.92 & 0.50 & 6.43 & 0.11\\
\tableline
\end{tabular}
\end{table}

\subsection{Initial conditions}
The initial conditions of each simulation are built as follows.
Working in Cartesian coordinates, we take a cube of edge
$\lzero=2\rzero$ centered in the origin, in which we generate a GRF
\begin{equation}
\delta(\xv)=\frac{1}{V}\sum_\kv \deltakv e^{\i \kv\cdot\xv }+\deltaminuskv e^{-\i \kv\cdot\xv},
\end{equation} 
where $\deltakv$ and $\deltaminuskv$ are independent random variables
with $\deltaminuskv=\deltakv^*$, $V=\lzero^3$ and the sum is performed
over half of the $\kv$-space (see \citealt{Bin08}, section 9.1.1). The
GRF $\delta(\xv)$ is fully characterized by its power spectrum
$P(\kv)=\langle|\deltakv|^2\rangle/V$, which we parameterize as
\begin{equation}
 P(\kv) =
  \begin{cases}
   \Pzero(k/\kzero)^{n} &\mbox{if}\quad \kzero \leq k\leq \kcutoff \\
   0       & \mbox{if}\quad  k < \kzero \quad\mbox{or}\quad k> \kcutoff. \\
  \end{cases}
\end{equation}
{Here $\Pzero$ and $n$ are, respectively, the power-spectrum amplitude
and index, $\kzero=2\pi/\rzero$ is the minimum wave-number and
$\kcutoff$ is the cut-off wave-number.}

We build a lattice of $\Nx\times\Ny\times\Nz$ points equally spaced in
$x$, $y$ and $z$.  At each lattice point we compute the density
field\footnote{We assume that the amplitudes of the fluctuations
  generated by the GRF are distributed log-normally because our initial
  conditions are meant to represent the non-linear phase of the
  collapse \citep{Kay01}.}  $\rho(x,y,z)=\rhobg\exp(\delta)$, where
$\rhobg(r)=\rhozero(r/\rzero)^{-\gamma}$ is an unperturbed background
power-law distribution ($r=\sqrt{x^2+y^2+z^2}$) and $\delta(x,y,z)$ is
the GRF defined above. Finally, $\rho$ is normalized to its maximum
value, so $0<\rho\leq 1$ independent of $\rhozero$.

Using the standard rejection technique, we generate a spherical
distribution of $N$ equal-mass particles with $r\leq\rzero$ and
density distribution $\rho$. The $x$, $y$ and $z$ coordinates of each
particle are then multiplied, respectively, by $\qx$, $\qy$ and $\qz$
to get a triaxial configuration. Velocity components $\vx$, $\vy$ and
$\vz$, extracted from a Gaussian distribution with vanishing mean and
unit variance, are temporarily assigned to each particle.  Once the
total potential energy $W$ and the temporary total kinetic energy $T$
are computed, all the velocity components are multiplied by
$\sqrt{\beta|W|/2T}$ in order to obtain a system with initial virial
ratio $\beta$.

\subsection{Parameters of the simulations}

The aim of our simulations is to isolate the effect on the end-product
of dissipationless collapse of the relative contribution of short- and
long-wavelength modes in the fluctuation power spectrum.  {Therefore,
  we present the results of a set of simulations differing only in the
  value of the power-spectrum index $n$, which spans the range $-3
  \leq n \leq 0$ (see Table~\ref{tab:par}).}  In all the simulations
the phases of the GRF modes are the same, as well as the other
parameters determining the initial conditions: $\Pzero=0.3$,
$\kcutoff=6\kzero$, $\gamma=1$, $\beta=0.01$, $\qx=6$, $\qy=4$,
$\qz=2$, and $\Nx=\Ny=\Nz=80$.  For comparison, we also ran a
simulation (named P0) with the same parameters as above, but no
fluctuations ($\Pzero=0$).

{The $N$-body simulations were run with the parallel collisionless
  code {\FVFPS} ({\FORTRAN} version of a fast Poisson solver;
  \citealt{Lon03,Nip03}), which is based on \citet{Deh02} algorithm,
  an efficient combination of the fast multiple method \citep{Gre87}
  and the tree code \citep{Bar86}.  The main parameters of the
  {\FVFPS} code are the number of particles $N$, the softening length
  $\epsilon$, below which the Newtonian force is smoothed, and the
  minimum value of the opening parameter $\theta_{\rm min}=0.5$, which
  determines the mass-dependent tolerance parameter $\theta$
  (analogous to the opening angle of \citealt{Bar86}) used to control
  the accuracy of the force approximation (see \citealt{Deh02} and
  \citealt{Lon03}). In our simulations we adopted $N=10^5$,
  $\theta_{\rm min}=0.5$ and $\varepsilon = 0.02 \rzero$.  The
  time-step $\Delta t$, which is the same for all particles, is
  allowed to vary adaptively in time as $\Delta t =0.3/(4\pi G
  \rho_{\rm max})^{1/2}$, where $\rho_{\rm max}$ is the maximum
  particle density.  Each simulation runs from $t=0$ to $t=100\tu$,
  where $\tu\equiv\sqrt{\rzero^3/GM}$ and $M$ is the total mass of the
  $N$-body system.  With this choice the system is fully virialized at
  the end of the simulation (the initial system's free-fall time is
  $\approx 10\tu$). The time-step values are in the range $8\times
  10^{-4}\lesssim \Delta t /\tu \lesssim 3\times 10^{-2}$.}

\begin{figure*}
\centerline{
\psfig{file=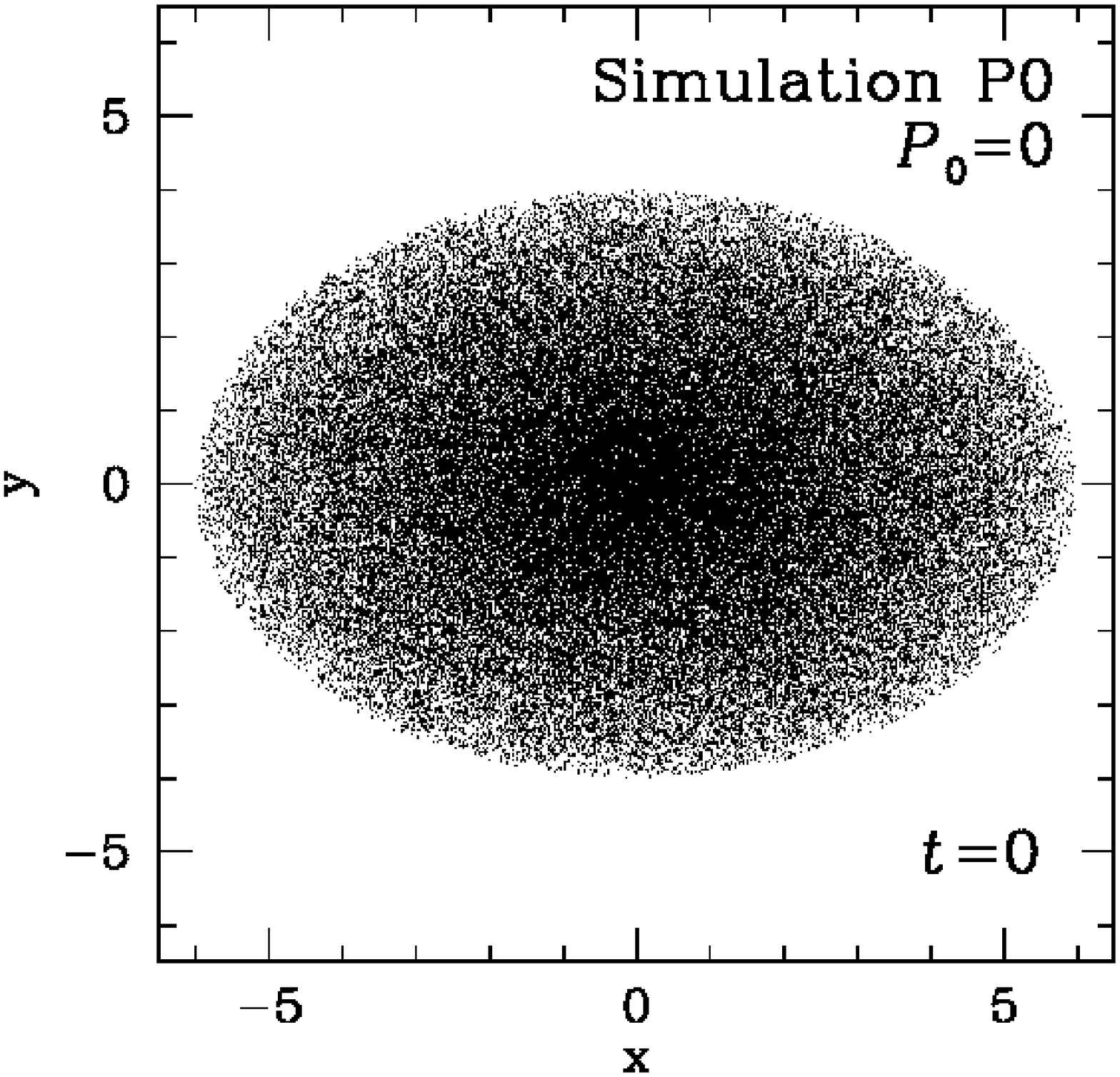,width=0.33\hsize}
\psfig{file=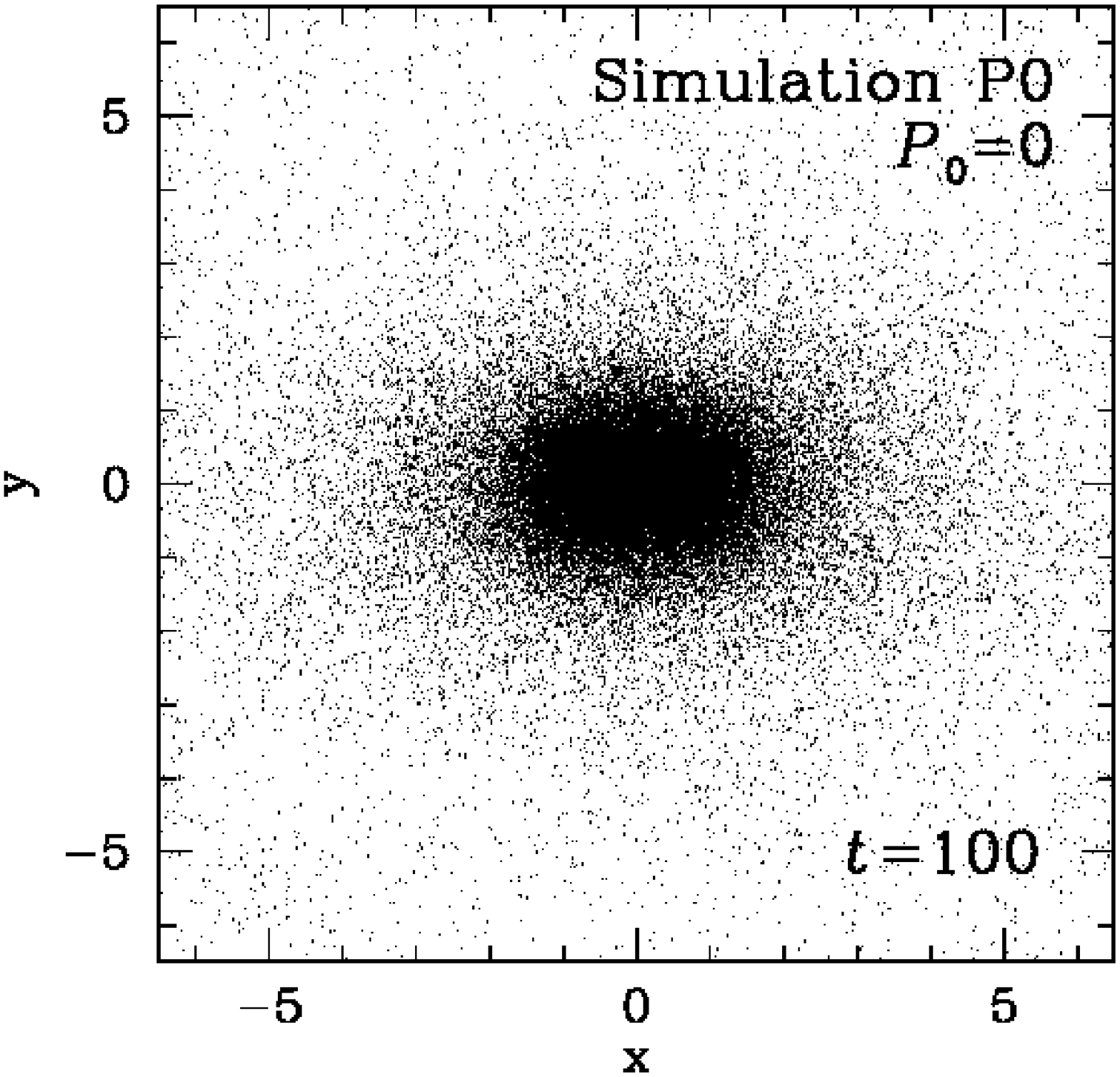,width=0.33\hsize}
\psfig{file=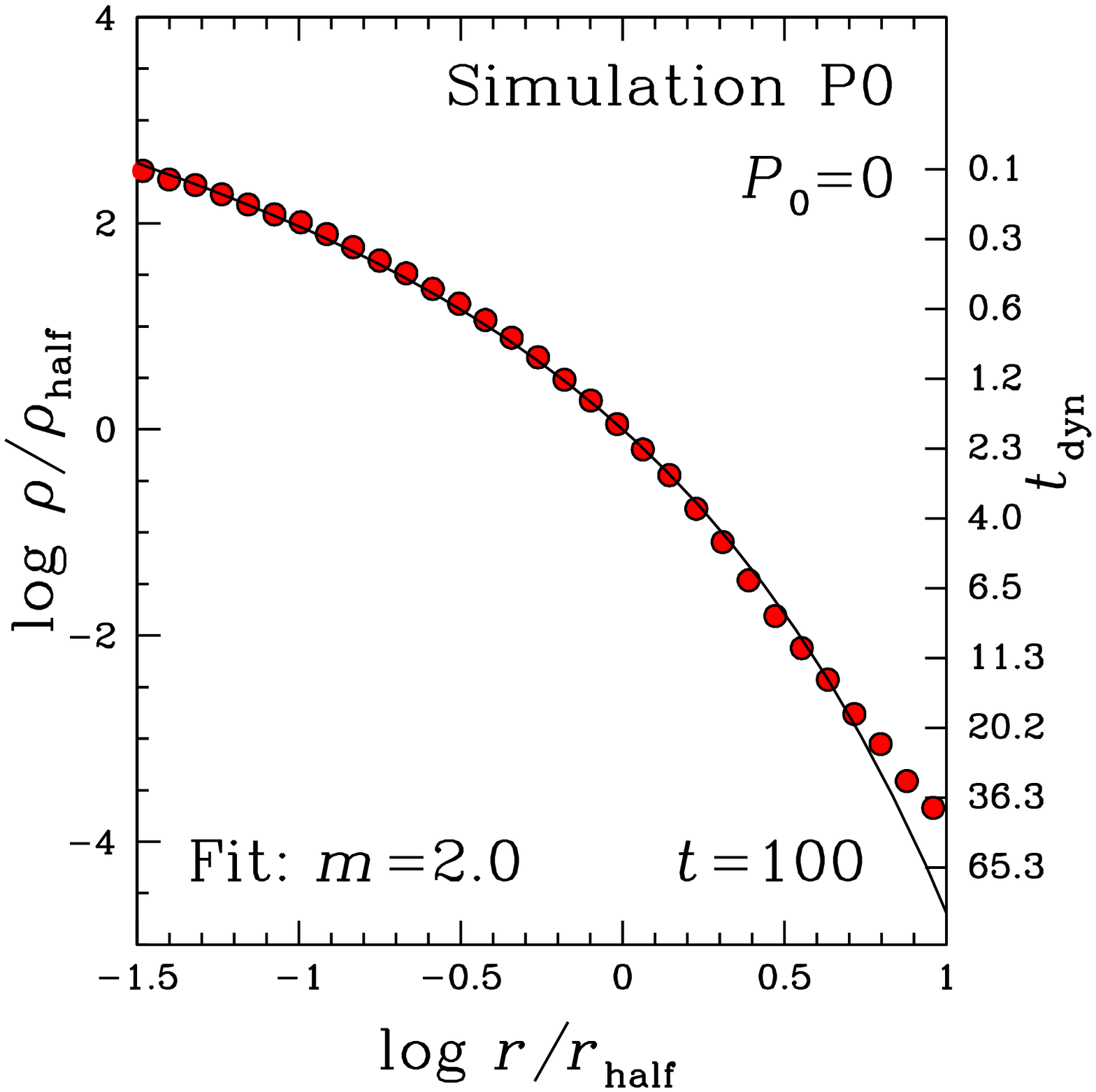,width=0.33\hsize}
}
\centerline{
\psfig{file=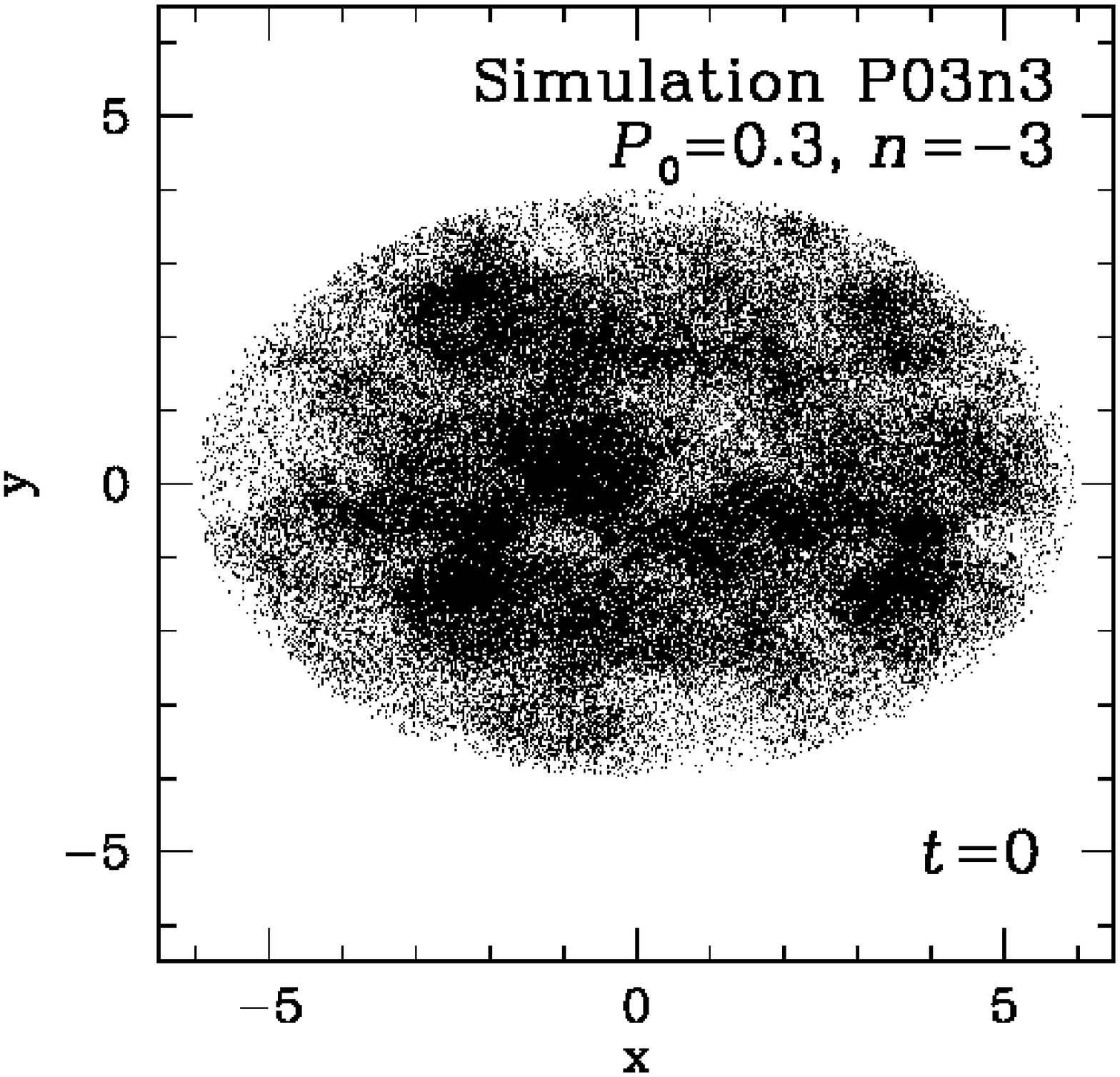,width=0.33\hsize}
\psfig{file=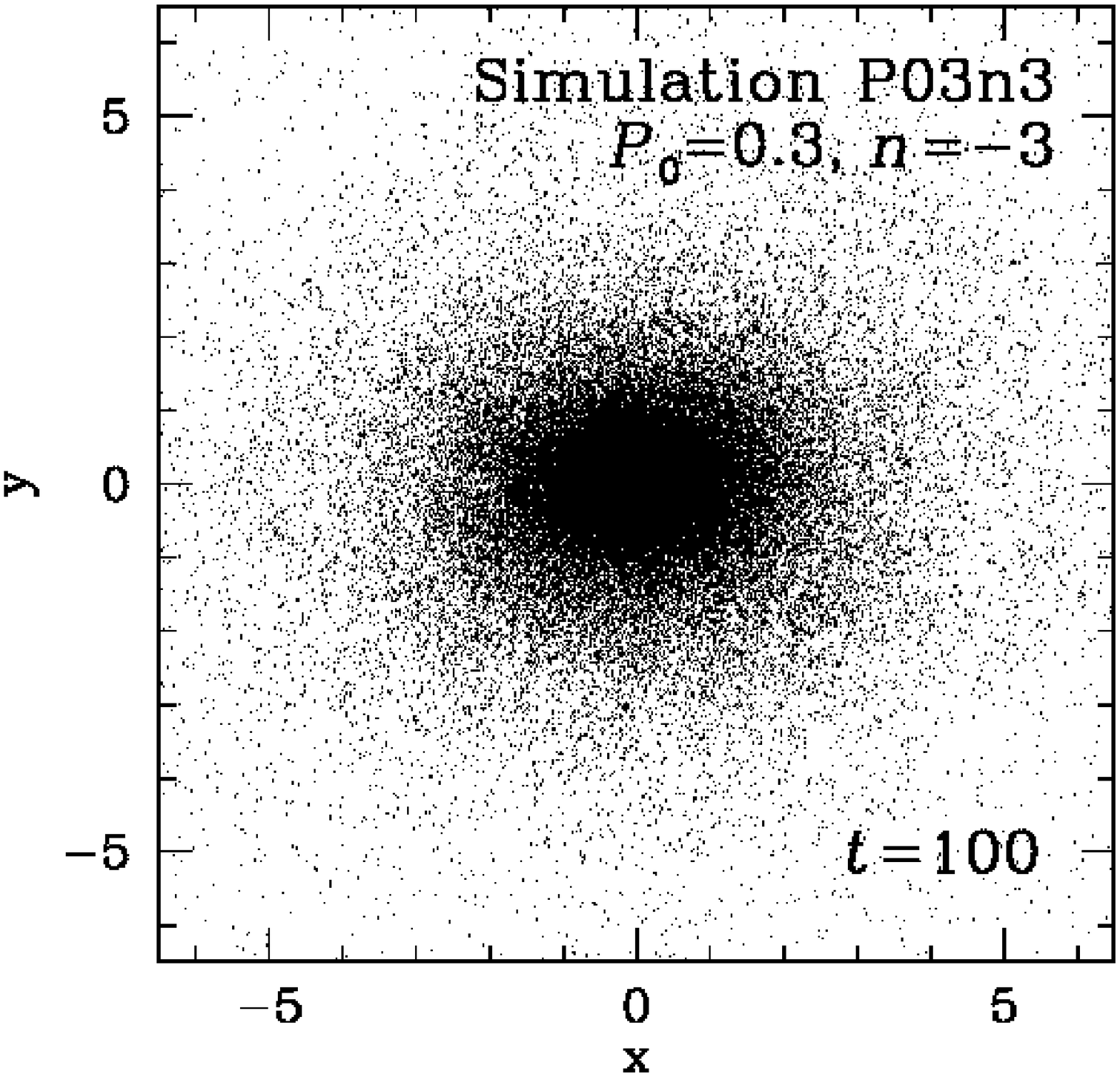,width=0.33\hsize}
\psfig{file=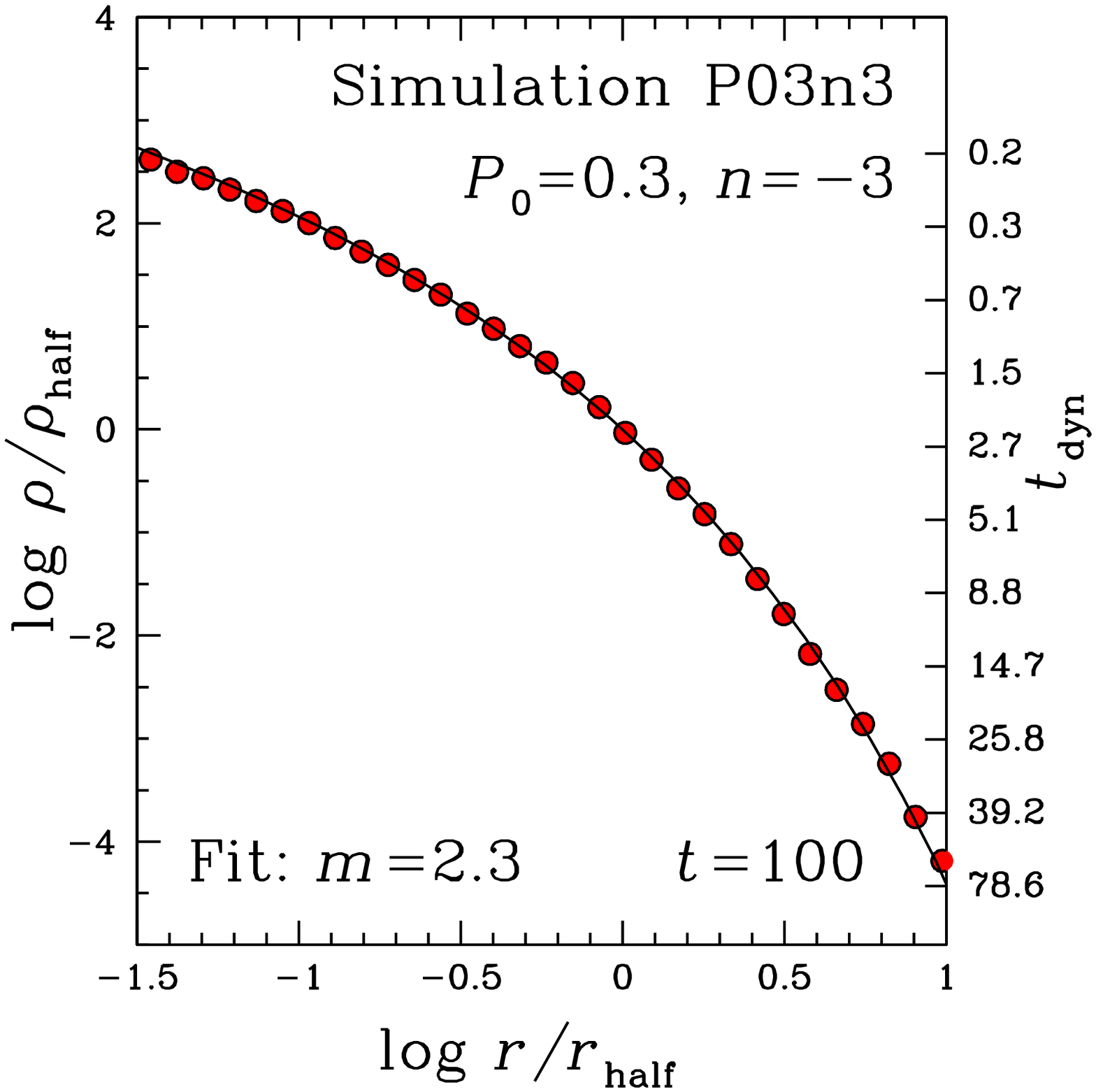,width=0.33\hsize}
}
\centerline{
\psfig{file=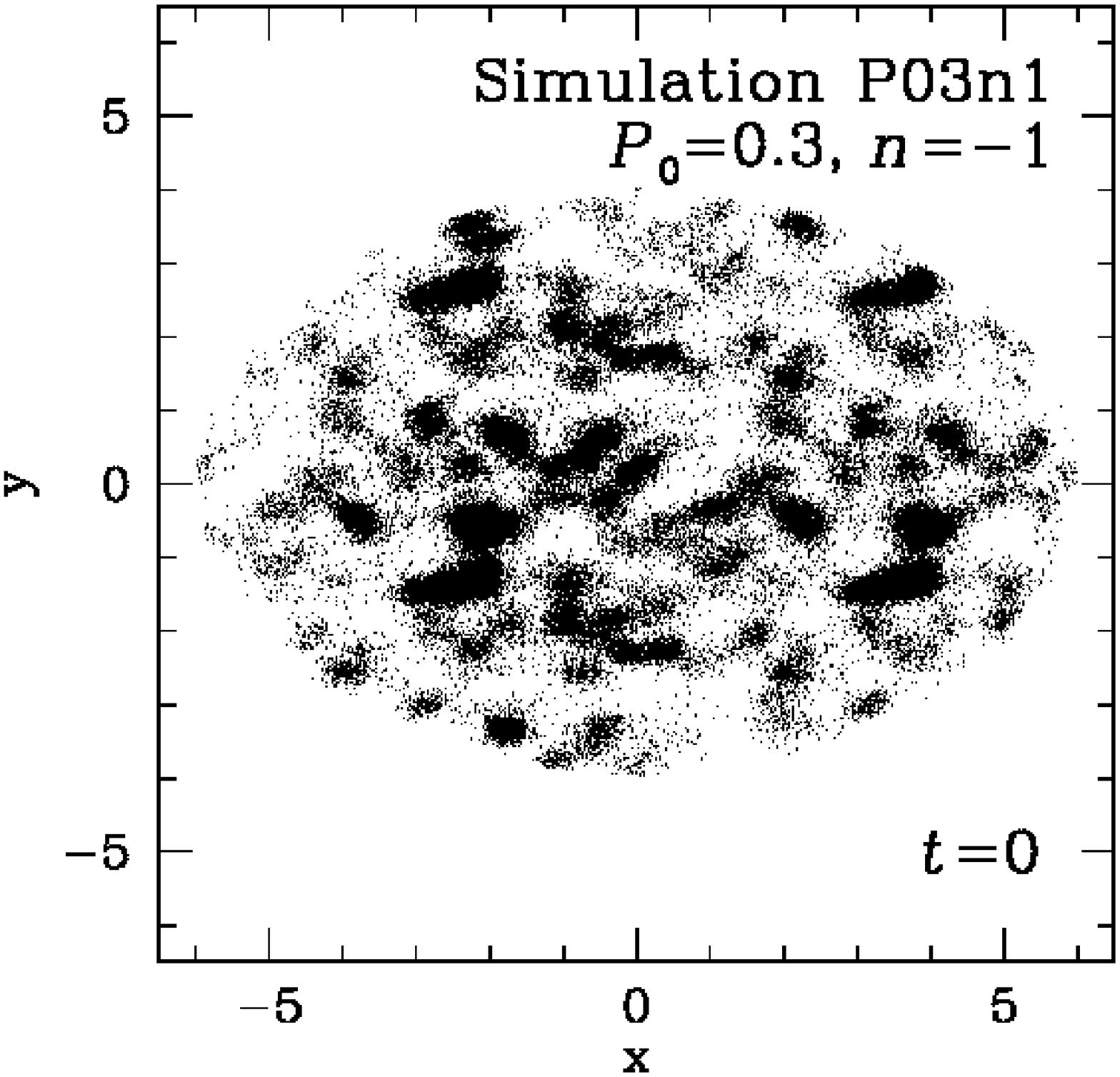,width=0.33\hsize}
\psfig{file=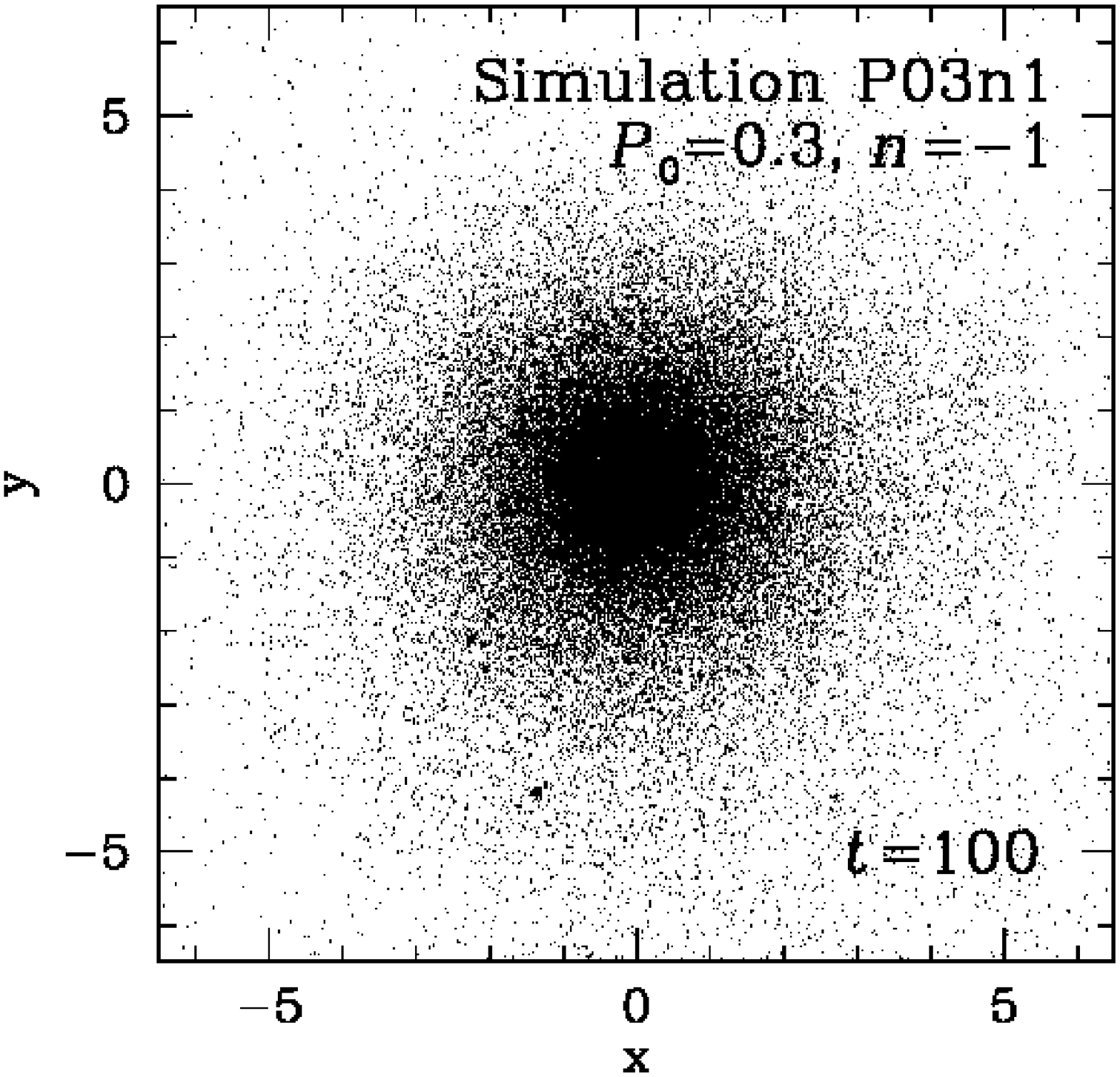,width=0.33\hsize}
\psfig{file=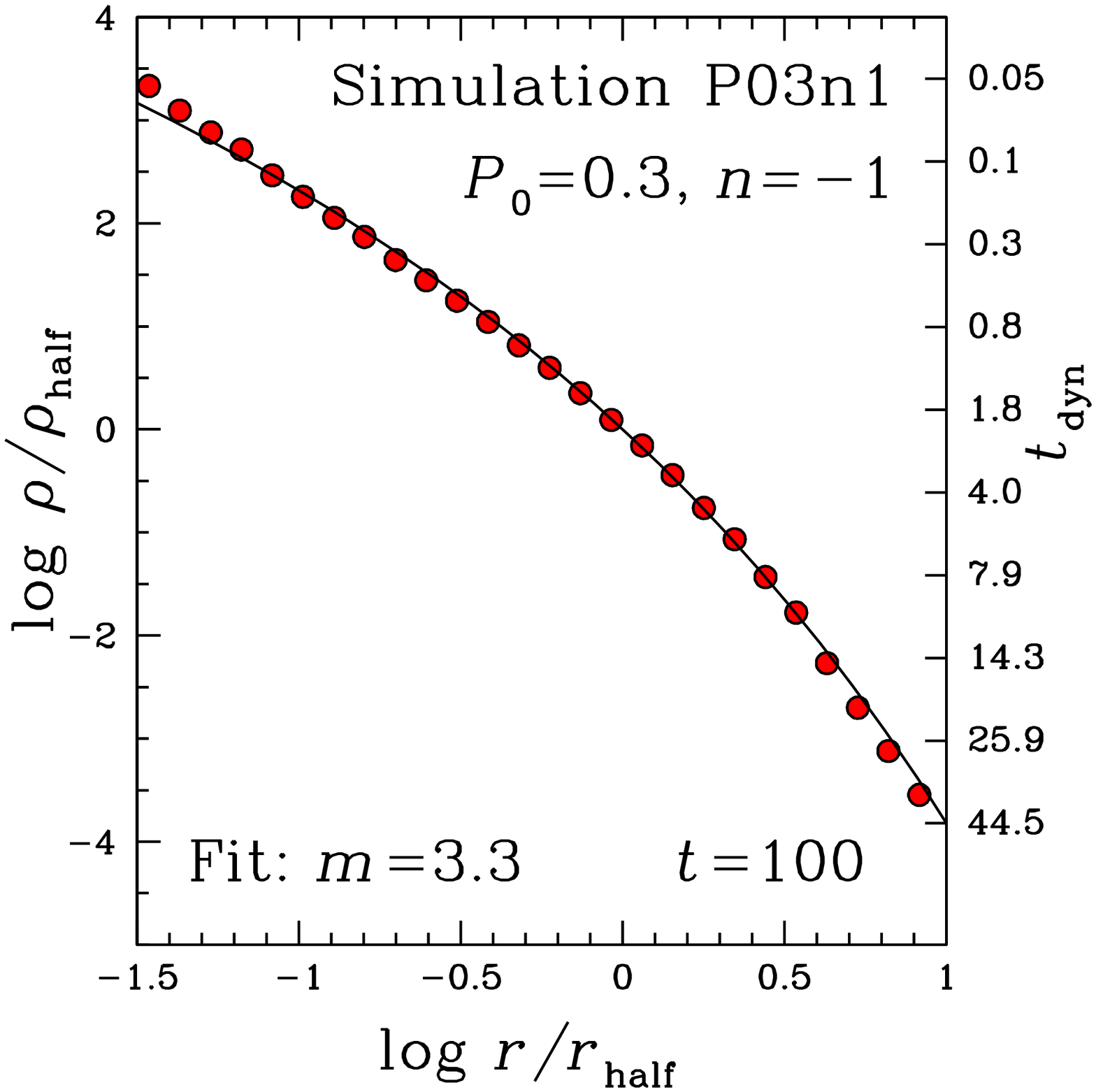,width=0.33\hsize}
}
\centerline{
\psfig{file=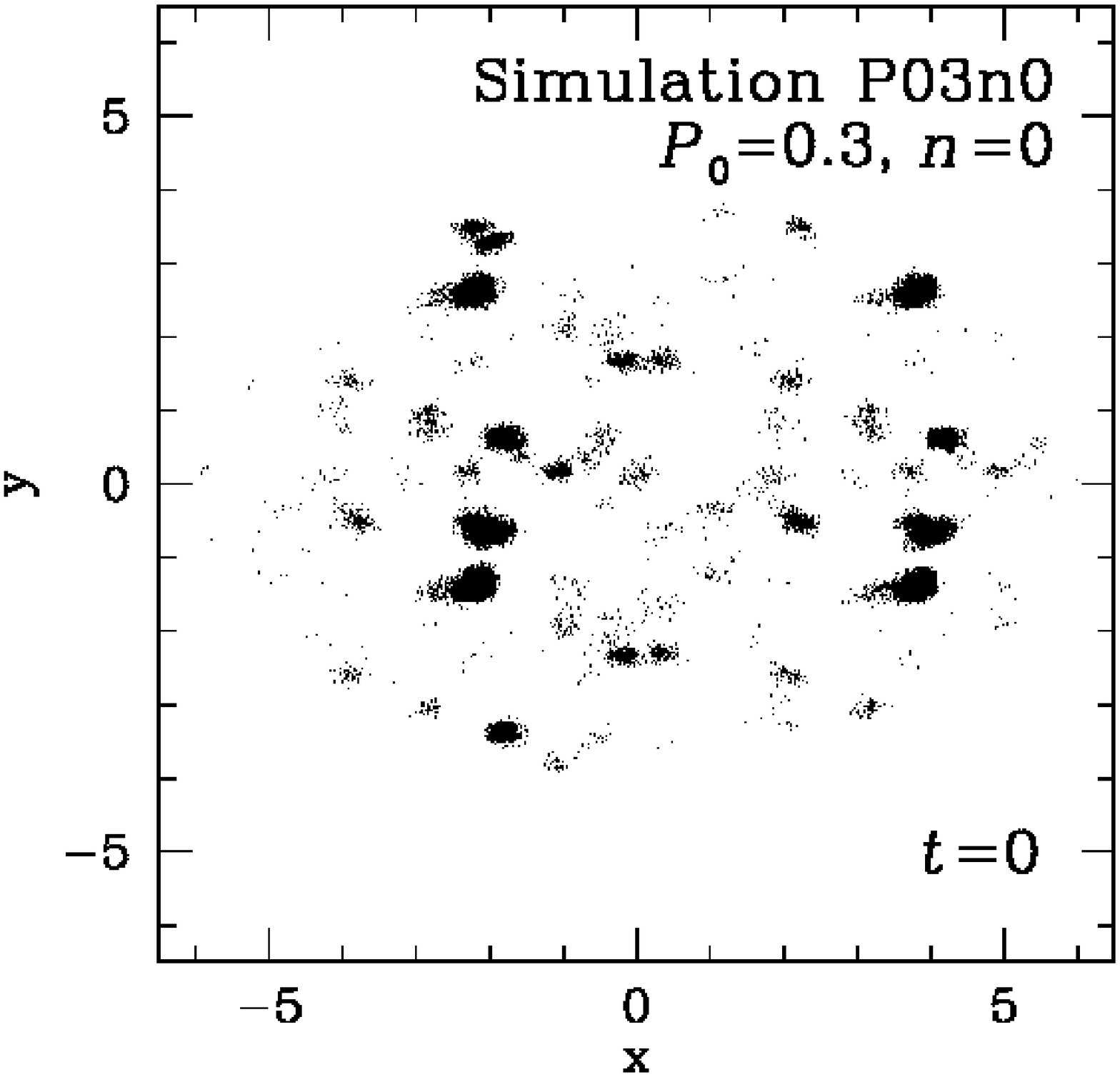,width=0.33\hsize}
\psfig{file=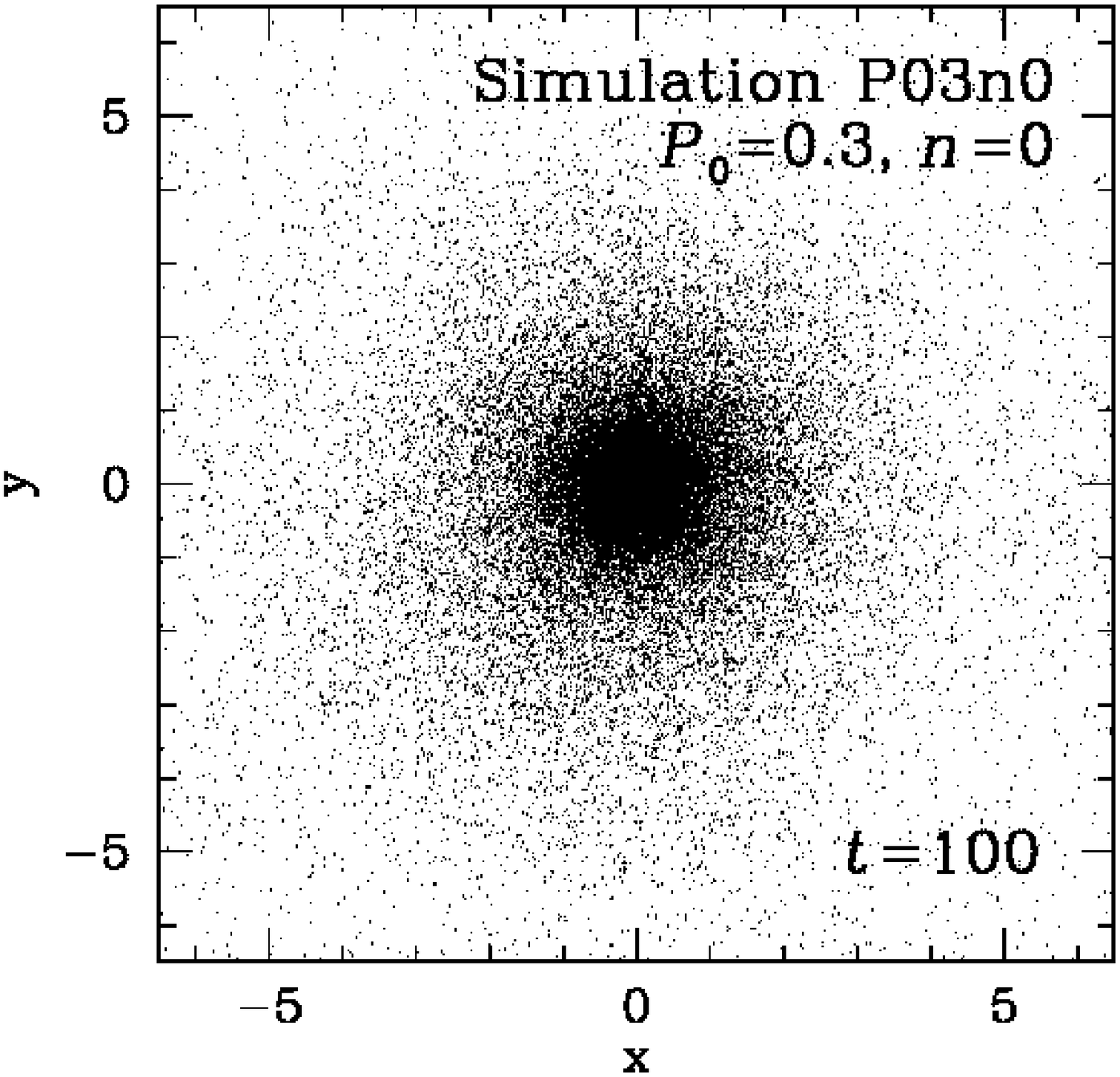,width=0.33\hsize}
\psfig{file=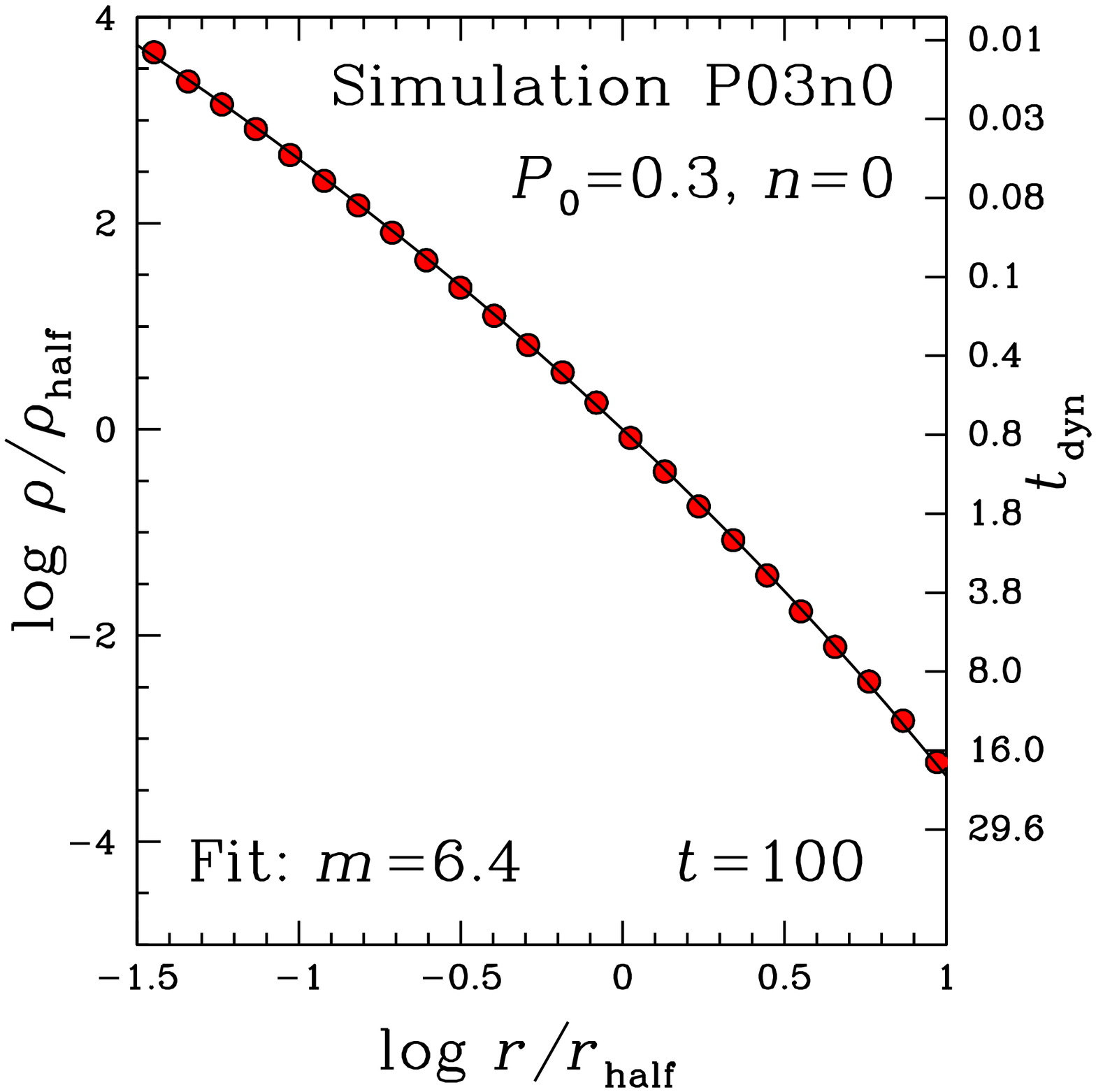,width=0.33\hsize}
}
\caption{Initial (left-hand column of panels) and final (central
  column of panels) particle distributions (projected along the
  shortest axis) of simulations P0, P03n3, P03n1 and P03n0 (from top
  to bottom). {The panels in the right-hand column show the
    corresponding final angle-averaged density profiles (circles)
    together with their best-fitting deprojected {\Sersic} profiles
    (solid curves); using the right-hand axes the circles can be
    interpreted as dynamical time $\tdyn$ as a function of radius for
    the final systems. Here times are in units of $\tu$, $x$ and $y$
    are in units of the scale radius $\rzero$, and $\rhohalf$ is the
    density at the half-mass radius $\rhalf$.}}
\label{fig:all}
\end{figure*}


\begin{figure}
\centerline{
\psfig{file=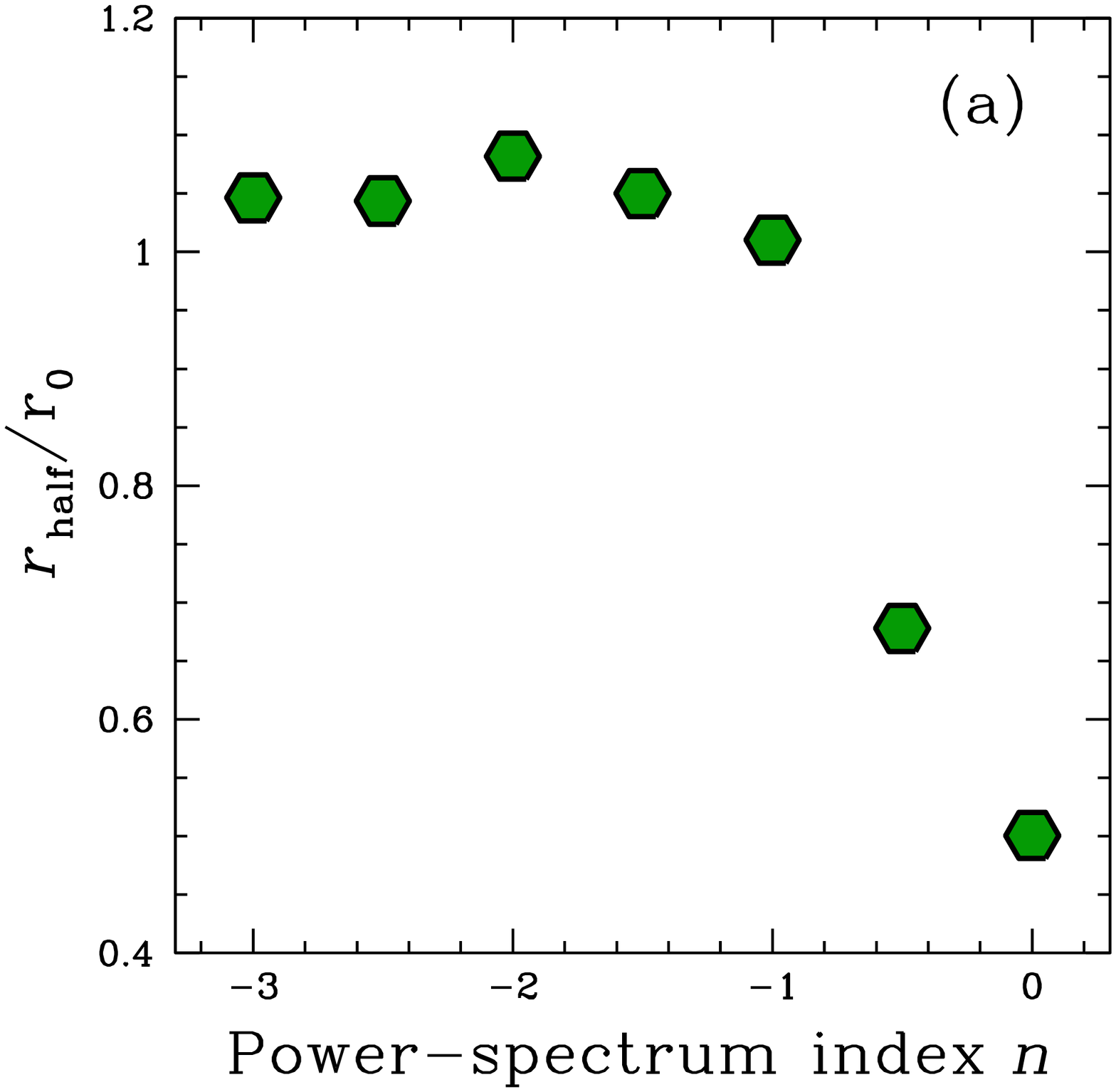,width=0.9\hsize}
}
\centerline{
\psfig{file=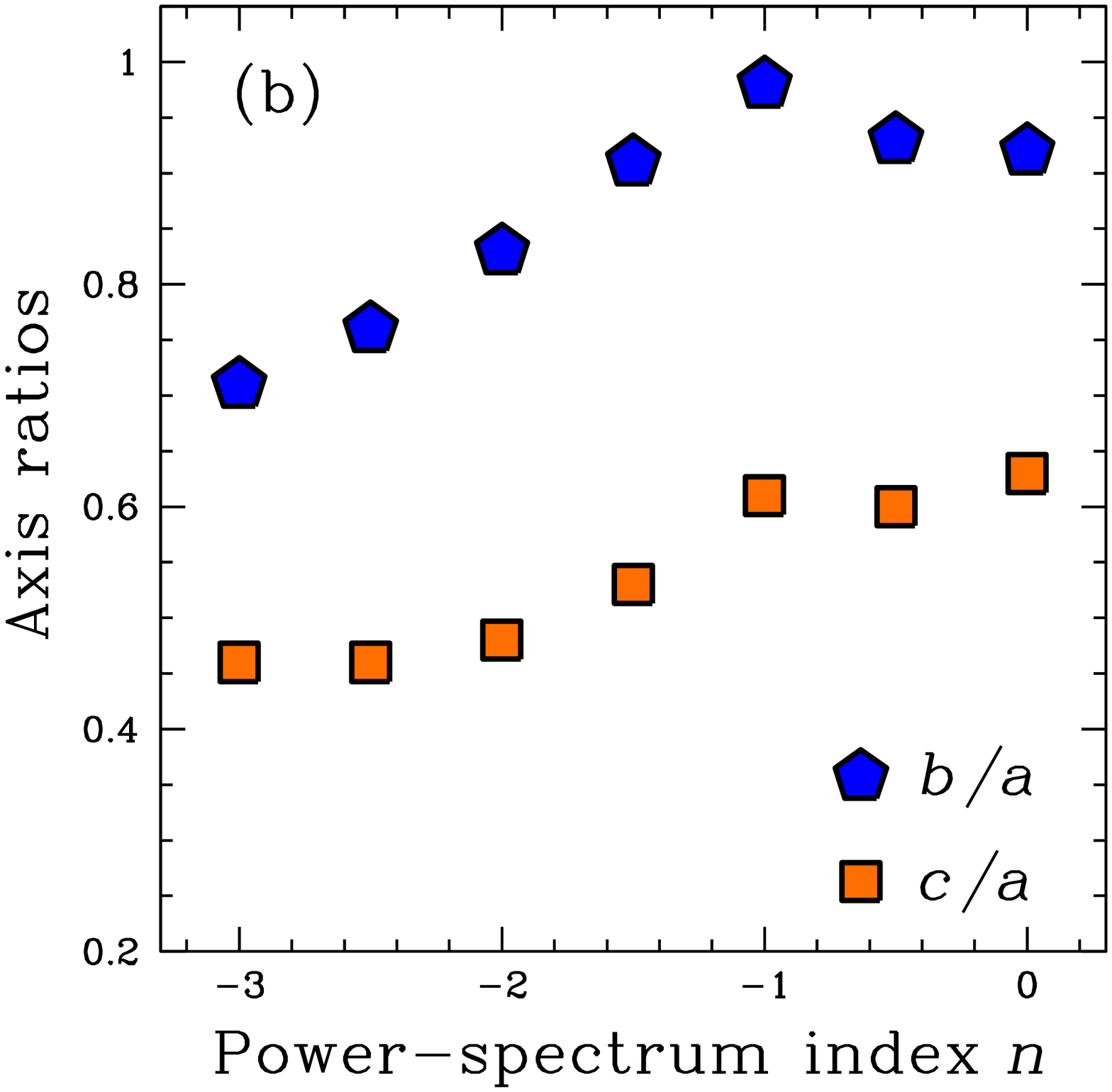,width=0.9\hsize}
}
\centerline{
\psfig{file=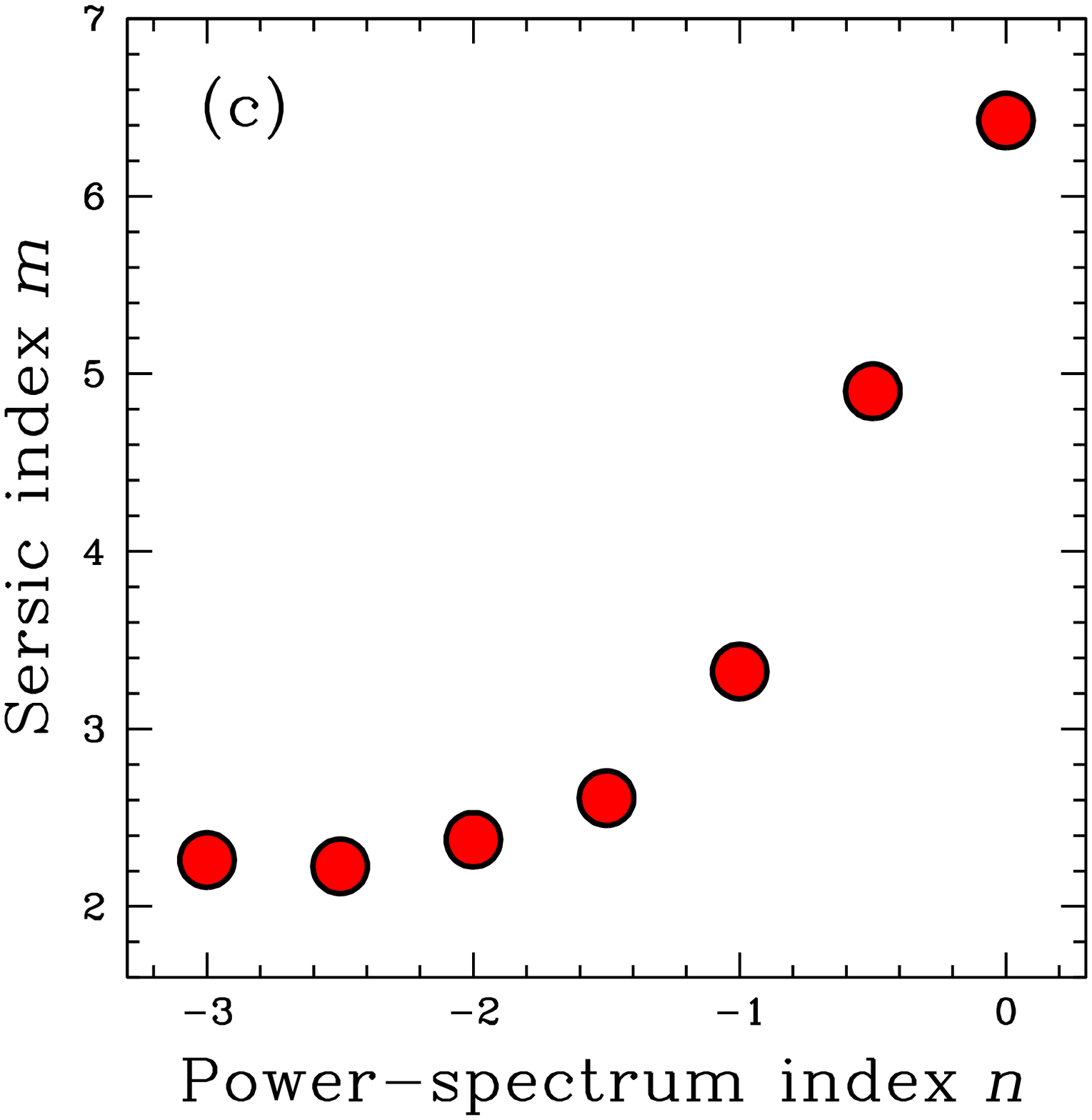,width=0.9\hsize}
}
\caption{{Final half-mass radius $\rhalf$ normalized to the scale
    radius $\rzero$ (panel a), axis ratios (panel b) and best-fitting
    {\Sersic} index $m$ (panel c) as functions of the initial
    fluctuation power-spectrum index $n$ of the $N$-body simulations
    P03n3, P03n25, P03n2, P03n15, P03n1, P03n05 and P03n0 (from left
    to right; see Table~\ref{tab:par}).}}
\label{fig:corr}
\end{figure}

\subsection{Results}

The intrinsic and projected properties of the collapse end-products
are determined as in \citet{Nip06}. At the end of the simulation only
bound particles are selected (the mass loss is in the range
2-11\%). The position of the center of the system is determined using
the iterative technique described by \citet{Pow03}.  Following
\citet{Nip02}, we measure the axis ratios $c/a$ and $b/a$ of the
inertia ellipsoid of the final density distribution, its
angle-averaged profile and half-mass radius $\rhalf$ ($a$, $b$ and $c$
are, respectively, the longest, intermediate and shortest axes).  

In Fig.~\ref{fig:all} we plot the initial and final particle
distributions (projected along the shortest axis), together with the
final angle-averaged density profiles of four representative
simulations: simulation P0 (with $\Pzero=0$) and simulations P03n3,
P03n1 and P03n0 (with $\Pzero=0.3$ and, respectively, $n=-3$, $n=-1$
and $n=0$).  As apparent from the plots in the left-hand column of
panels, the initial distribution becomes more and more clumpy from top
(simulation P0, with homogeneous initial conditions) to bottom
(simulation P03n0, in which most particles are initially in relatively
few small clumps). The corresponding end-products (central column of
panels of Fig.~\ref{fig:all}) tend to be more extended when the
initial conditions are more homogeneous (upper plots) and more compact
when the initial conditions are clumpier (lower plots). {The final
  half-mass radius $\rhalf$ ranges from $\approx \rzero$ when $n=-3$
  down to $\approx 0.5 \rzero$ when $n=0$ (see Table~\ref{tab:par} and
  Fig.~\ref{fig:corr}a). The end-products are typically triaxial with
  $0.43\lesssim c/a\lesssim 0.63$ and $0.56\lesssim b/a\lesssim
  0.98$.}  The trend with $n$ is that the systems tend to be almost
prolate for smooth initial conditions and almost oblate for clumpy
initial conditions (see Table~\ref{tab:par} and Fig.~\ref{fig:corr}b).

{The panels in the right-hand column of Fig.~\ref{fig:all} show
  the final angle-averaged density profiles of the aforementioned
  representative simulations for $-1.5\leq \log (r/\rhalf) \leq 1$.
  Over this radial range the dynamical time\footnote{{At each
      angle-averaged radius $r$ we define the dynamical time as
      $\tdyn(r)=\sqrt{3\pi/16 G\rhobar(r)}$, where
      $\rhobar(r)=3M(r)/4\pi r^3$ is the average density within $r$
      and $M(r)$ is the mass contained within $r$.}}  $\tdyn(r)$ is
  shorter than the simulation time-span $100\tu$ (see rightmost axes
  in Fig.~\ref{fig:all}), so the density profiles can be considered
  stationary. From these plots it is apparent that, consistent with
  the idea of \citet{Cen14}, the final angle-averaged density profile
  is steeper in the center and shallower in the outskirts when the
  initial conditions are clumpier.} We quantify this finding by
comparing the final distributions with the {\Sersic} law
(equation~\ref{eq:ser}).  {Under the assumption of spherical
  symmetry and position-independent mass-to-light ratio, the intrinsic
  stellar mass density distribution corresponding to
  equation~(\ref{eq:ser}) can be obtained in integral form
  \citep[][]{Cio91}.}  A simple approximation of the deprojected
{\Sersic} profile, which we adopt in this work, is
\begin{equation}
\rho(r)=\rhohalf\left(\frac{r}{\rhalf}\right)^{-p}\exp{\left[\left(\frac{\rhalf}{\rs}\right)^{\nu}-\left(\frac{r}{\rs}\right)^{\nu}\right]}
\label{eq:deproj}
\end{equation}
\citep{Lim99}, where $\nu=1/m$, $p=1-0.6097\nu+ 0.05463\nu^2$,
$\rhohalf\equiv\rho(\rhalf)$ is the density at the half-mass radius
and $\rs$ is a characteristic radius related to $\rhalf$ by
\begin{eqnarray}
&&\ln\left(\frac{\rhalf}{\rs}\right)=\ln\left(1.356-0.0293\nu+0.0023\nu^2\right)\nonumber\\
&&+\frac{0.6950-\ln(\nu)}{\nu}-0.1789.
\end{eqnarray}
The final angle-averaged density profiles of our $N$-body simulations
are very well represented by the deprojected {\Sersic} law
(equation~\ref{eq:deproj}). The profiles are fitted taking $m$ as only
free parameter, because $\rhalf$ and $\rhohalf$ are fixed by the
measured values. The fits performed over the radial range $0.04\leq
r/\rhalf\leq 10$ give values of the {\Sersic} index in the interval
$2\lesssim m \lesssim 6.5$ with small associated uncertainties
$0.02\lesssim \sigma_m\lesssim 0.15$ (see Table~\ref{tab:par} and
panels in the right-hand column of Fig.~\ref{fig:all}). {In
  simulation P0 (with smooth initial conditions) we perform the fit
  over the smaller radial range $0.04\leq r/\rhalf\leq 5$, because the
  profile has a power-law tail at large radii (see top-right panel of
  Fig.~\ref{fig:all}), which is reminiscent of the core-halo
  structure, a well-known feature of collapses starting from
  homogeneous initial conditions \citep{Lon91}.}  Fig.~\ref{fig:corr}c
shows that the best-fitting {\Sersic} index $m$ increases for increasing
$n$: clumpier initial conditions lead to higher values of $m$.

\section{Discussion and conclusions}

The results of our simulations confirm the conjecture of
\citet{Cen14}: the density profile of dissipationless collapse is
steeper in the center and shallower in the outer parts if the
fluctuation power spectrum of the initial conditions is dominated by
short-wavelength modes. Vice versa, power spectra dominated by
long-wavelength fluctuations lead to density profiles that are shallow
in the center and steep in the outskirts. The end-products of our
simulations have density distributions well represented by the
deprojected {\Sersic} law with index in the range $2\lesssim m \lesssim
6.5$. For increasing spectral index $n$ the best-fitting {\Sersic} index
$m$ increases, the half-mass radius $\rhalf$ decreases, and the
systems tend to move from almost prolate to almost oblate intrinsic
shape.

Of course, the exact values of the measured quantities are expected to
depend on the details of the initial conditions: for instance, while
here we find $m\simeq 2$ for the end-product of simulation P0 (with
smooth initial $\rho\propto r^{-1}$ density profile), it is well known
that the end-product of a cold collapse with smooth initial
\citet{Plu11} density distribution is extremely well fitted by the
\citet{dev48} $m=4$ profile (\citealt{Lon91,Nip06}). Therefore, the
above range $2\lesssim m \lesssim 6.5$ must not be taken at face
value. {However, it is interesting to notice that, based on the
  results of the present work and of previous studies, it appears hard
  to get $m<2$ with purely dissipationless processes, consistent with
  the expectation that the formation of $m\approx 1$ systems
  (typically disks) requires dissipative processes.}

It is interesting to compare our results with those of previous
similar investigations. A very interesting work is the paper of
\citetalias{Kat91}, who attempted a systematic study of the effect of
GRF power spectrum on the structure of virialized systems. Though the
initial conditions of \citetalias{Kat91} simulations, which were meant
to represent conditions before turn-around, are different from ours in
many respects, based on the results of the present work we should
expect that the final density profiles of \citetalias{Kat91} depend on
the initial fluctuation power spectrum.  In fact, the conclusion of
\citetalias{Kat91} is that the final profiles do not depend
significantly on the power-spectrum slope: however, when compared to
today's standard, the resolution of the simulations of
\citetalias{Kat91} is rather poor ($\approx 4000$ particles), so it is
likely that detailed differences in the density profiles were obscured
by numerical noise.

A set-up in a sense more similar to ours was that of
\citetalias{Agu90}, who did not included fluctuations in their initial
conditions, but considered the dissipationless collapse of smooth
triaxial particle distribution with the same initial density field as
our background distribution ($\rhobg\propto r^{-1}$). Our simulation
P0 (with no fluctuation; $\Pzero=0$) is therefore very similar to
those of \citetalias{Agu90} with virial ratio $\beta\approx 10^{-2}$
and actually, consistent \citetalias{Agu90}, we find a prolate final
system with axis ratios $c/a\sim b/a\approx 1/2$. However, while we
find best-fitting {\Sersic} index $m\simeq 2$, \citetalias{Agu90} report
that their final distributions are well fitted by $m=4$. Again, this
is likely a matter of resolution: \citetalias{Agu90}, with 5000
particles, can follow the profile down to $\approx 0.5\rhalf$, where
the difference between the $m=2$ and $m=4$ profiles is hard to detect,
while in this work we have been able to fit the profiles down to
$0.04\rhalf$.

Independent support to Cen's model and to our results comes from
numerical studies of dissipationless galaxy mergers.  A galaxy growing
in a region of the Universe dominated by fluctuations on small scales
is expected to form by several mergers of smaller subunits. {The
  finding that the best-fitting {\Sersic} index $m$ increases for
  increasing fluctuation power-spectrum index $n$ is therefore
  consistent with the results of numerical simulations showing that
  dissipationless mergers make the {\Sersic} index increase
  \citep{Nip03}.  In particular, the dissipationless accretion of
  small satellites is believed to be the most promising mechanism to
  form high-$m$ systems \citep{Hil13}. However, dissipative processes
  can also contribute to raise $m$, as found for instance in
  simulations of mergers between gas-rich disk galaxies
  \citep[e.g.][]{Hop09}.}

In this work we have provided quantitative support to the idea that
the origin of the {\Sersic} law is related to the fluctuation power
spectrum in the initial conditions of galaxy formation. Still, our
results are based on toy models that neglect all the complexities of
proper galaxy formation theories. In the future it will be interesting
to explore this idea more realistically by using cosmological
simulations with distinct baryonic and dark matter components, and
including the all-important dissipative processes.

\acknowledgments

I am grateful to an anonymous referee for useful suggestions that
helped improve this Letter. I acknowledge financial support from PRIN
MIUR 2010-2011, project ``The Chemical and Dynamical Evolution of the
Milky Way and Local Group Galaxies'', prot. 2010LY5N2T.

\end{document}